# Calibrated steganalysis of mp3stego in multi-encoder scenario

Hamzeh Ghasemzadeh

*Abstract—* **Comparing popularity of mp3 and wave with the amount of works published on each of them shows mp3 steganalysis has not found adequate attention. Furthermore, investigating existing works on mp3 steganalysis shows that a major factor has been overlooked. Experimenting with different mp3 encoders shows there are subtle differences in their outputs. This shows that mp3 standard has been implemented in dissimilar fashions, which in turn could degrade performance of steganalysis if it is not addressed properly. Additionally, calibration is a powerful technique which has not found its true potential for mp3 steganalysis. This paper tries to fill these gaps. First, we present our analysis on different encoders and show they can be classified quite accurately with only four features. Then, we propose a new set of calibrated features based on quantization step. To that end, we show quantization step is a band limited signal and steganography noise affects its high frequency components more prominently. By applying a low pass filter on quantization steps, we arrive at an estimation of quantization step, which in turn is used for calibrating the features.**

## I. INTRODUCTION

Due to recent advances in telecommunications, wireless technology, and wide spread usage of social media, computers and internet have become an integrated part of our everyday life. Nowadays, people can perform a wide variety of tasks including banking, storing/sharing information, connecting with their family and friends and etc. only using internet and their cellphones. This new era has brought major conveniences with itself, but on the other hand, if security of these services are compromised the damage could be catastrophic. To prevent such circumstances different security mechanisms have been introduced. For example, if encryption is used properly it could provide confidentiality [1]. But, preventing traffic analysis is something that encryption cannot achieve [2]. That is, although adversary cannot read encrypted messages, but there are lots of other valuable information that still he can acquire. For example, the mere exchange of encrypted message reveals value of its content. Also, the adversary can see sender and receiver, frequency and patterns of communications, end etc. Steganography can provide the means for hiding messages inside a carrier without arising any suspicion at all and therefore solve these problems by adding an extra layer of protection.

Availability of camera and sound recorder in most cellphones with wide usage of multimedia signals in social media make multimedia signals perfect candidates for steganography. But at the same time there are some concerns regarding usage of steganography including determining security of steganography methods and preventing improper usage of steganography. Steganalysis is the science of breaking steganography methods and its aim is to address these concerns. Reviewing literature on audio steganalysis shows they can be categorized into uncompressed (wave) and compressed (mp3) methods [3].

In the uncompressed domain one of the first work was [4] where deviation between signal and its denoised version was captured using audio quality metrics. Another work used chaotic-based measurements for speech steganalysis [5]. Later, Mel frequency cepstrum coefficients (MFCC) which are based on ear were used for steganalysis [6]. Ghasemzadeh et al. argued that using features that are based on model of ears are counter intuitive and they proposed a new set of features based on a model that was maximally deviant from ear [7]. Recently the same authors showed that LSB re-embedding can be used as a universal calibration method for audio steganalysis [8] and it can achieve very good results. Interested readers may refer to [3] for a more comprehensive review of audio steganalysis methods.

In the compressed domain Westfeld did the pioneering job [9]. They noticed, mp3sego would produce files with lower bit rate if bit rate of mp3 is not held constant. They also showed std of block length can capture deviation between cover and stego even in constant bit rate mp3s. Other researchers noticed that mp3sego disturbs distribution of quantization step of mp3 [10] and those deviations are amplified if second order derivative of quantization step is used. A different approach was pursued in [11] where statistical characteristic of bit reservoir was employed. The work also used recompression calibration for improving the results.

Due to the mechanism of data hiding in mp3stego, distributions of modified discrete cosine transform (MDCT) coefficients change. To capture these anomalies different approaches have been proposed. Generalized Gaussian

H. Ghasemzadeh is with Michigan State University, MI, USA (e-mail: ghasemza@msu.edu).

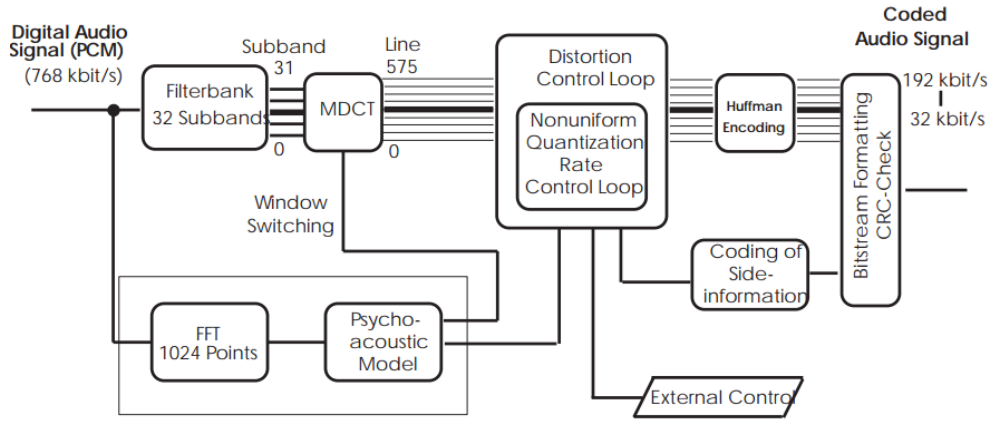

Fig 1. Diagram of mp3 Encoder [16]

distribution (GGD), moments of MDCTs in different subbands, and markov transition probabilities of inter and intra frames were used in [12]. The work investigated different feature selections and showed that a two level approach based on anova and support vector machine-recursive feature elimination (SVM-RFE) achieves the highest accuracy. Finally, Jin et al. showed that if markov features are extracted from difference of absolute values of MDCT across different channels considerable gain is achieved [13].

Comparing popularity of mp3 and wave over internet with the amount of works published on each of them shows that mp3 steganalysis has not found adequate attention. More importantly, investigating existing mp3 steganalysis methods shows that a major factor has been overlooked, which could degrade performance of existing methods considerably. Different implementations of mp3 standard exist and in fact, our analysis showed they have quite dissimilar characteristics. Also, we noticed that the differences are present in every aspect of mp3 standard. That is, different implementations have dissimilar statistical behaviors in their block length, quantization step, bit reservoir, and even their MDCTs. Unfortunately, previous works have not considered this factor and have only used one encoder for their evaluations. Additionally, it is known that calibration can lead to major improvements for steganalysis [8, 14]. Yet, its true potential for mp3 steganalysis has not been used. This paper tries to fill these gaps. First, we present our analysis on different encoders and show they can be classified quite accurately. Then, we propose a set of calibrated features based on quantization step. To that end, we show quantization step is a band limited signal and steganography noise has a more prominent effect on its high frequency components. By applying a low pass filter on quantization steps, we arrive at an estimation which in turn is used for calibrating the features.

The rest of this paper is organized as follows. Section 2 presents some preliminaries on mp3 standard and how mp3stego works. Section 3, is devoted to analysis of different encoders and the proposed calibrated features. Experimental results are presented in section 4. Finally, after some discussions, conclusions are drawn.

## II. PRELIMINARIES

*A. Mp3 compression algorithm:*

MP3 is a lossy format that provides good compression rate with excellent quality. The algorithm uses different techniques to achieve this. First, it uses model of human auditory system for finding portions of signal that are oblivious to humans' ears. Then, it uses non uniform quantization and entropy coding for efficient conversion of signal into bit stream. Finally, it uses a bit reservoir to save extra bits from simple frames and later uses them for more complex ones while maintaining constant bit rate [15].

Compression algorithm takes one frame of 1152 samples at a time and extract its spectral and psychoacoustic characteristic. This information is used for controlling parameters and maintaining inaudible distortions. The frame is then split into two granules and they are converted into frequency domain through polyphase filter and MDCT. Then, a two-level nested loop is employed for allocating bits to each granule. The inner loop finds suitable quantization step that meets with available bit budget and outer loop makes sure that compression noise is inaudible. Finally, quantized MDCTs and the data necessary for their decoding are muxed and written in the output bit stream. Fig. 1 shows these steps [16].

*B. Mp3 bit stream and its analysis*

In what follows, italic words are used in accordance with notations of mp3 standard.

*Main_data_begin*: bit reservoir allows MDCTs to be sent before their corresponding SI. The value of this offset is called *main_data_begin* and it is part of SI.

*Scalefactor Selection Information (scfsi)*: MDCTs are divided into a set of critical bands called *scalefactors*. Also, *scalefactors* are divided into four groups and if spectra of one group does not change in two granules, *scalefactors* are transmitted once and the extra bits are used for achieving better quality. For each group one bit of flag is sent in SI to determine if this technique was used.

*Big_values*, *region0_count* and *region1_count*: MDCTs of each granule are divided into three main regions of *big_values*,

*count1*, and *rzero*. Furthermore, *big_values* is divided into three sub-regions called *region0*, *region1*, and *region2*. Information regarding these regions are sent in SI. For example, *region0_count* and *region1_count* determine the number of scalefactors in each of these regions.

*Preflag*: if compression noise at high frequencies is audible, high frequency regions are modified according to a preemphasis table. *Preflag* determines if this strategy was used.

*Block_type*: mp3 standard defines a set of windows for providing different tradeoffs between time and frequency resolution. Typically, the long window is used but if there are considerable deviations between spectrums of two frames, short window is used. The standard also defines two other windows for transition between short and long windows. SI contains information about the type of window that was used for each granule.

*Global_gain* is part of SI and it determines quantization step.

*Table_select*: after quantization, MDCTs are Huffman coded. Mp3 defines different tables for this purpose. Index of selected table is part of SI.

*Part2_3_length* determines the number of bits that were used for storing values of scalefactors and quantized MDCTs in each granule.

### C. Data hiding in mp3stego

Mp3stego is an open source data hiding algorithm that is built on top of 8Hz encoder. The algorithm uses SHA-1 hash function for randomly selecting a set of frames and then embeds bits of message as parity of their *part2_3_length* field. Embedding algorithm works directly on uncompressed samples of cover and it embeds the message during compression itself. To that end, the algorithm adds a second criterion to inner loop of mp3 compression. So, not only should the existing bit budget be enough for encoding a granule, but also parity of its *part2_3_length* should match with message. Therefore, if parity of *part2_3_length* does not agree with the message, the inner loop is executed again [17].

## III. THE PROPOSED METHOD

In this section we show MDCTs, *part2_3_length* and many other important fields of SI which previously have been used for steganalysis, have dissimilar distributions for different encoders. Therefore, if covers are from multiple encoders (which is the realistic case), decision criteria between cover and stego would change and performance of system could diminish. In this section we propose a novel set of features for encoder classification. Then, we present a set of calibrated features for steganalysis of mp3stego.

### A. Encoders Classification

Our initial investigation showed there are some deviations between outputs of different encoders. These deviations stem from different sources including incorrect implementation of the standard, trading optimality for speed, and lack of details in some aspects of the standard [18]. Also, the same work showed that these subtle differences could be exploited for differentiating between different encoders. This section investigates deviations that exist between popular mp3 encoders. For this purpose, ten encoders including 8Hz, adobe audition, blade, fastenc, gogo, jetaudio, L3enc, lame, plugger, and xing (from audio catalyst) were used.

*1) Analysis of SI*

Investigating encoders showed they assign bit budget in dissimilar fashion to different regions of spectrum. This, results in different distributions of *big_values*, *region0_count* and *region1_count* in SI. Fig 2 shows these differences.

SI of every frame contains four fields of scfsi. Our investigation showed only lame and xing use this feature and these flags are always zero for other encoders. Furthermore, lame and xing uses this feature quite differently. Preflag is another parameter that was implemented differently. For example, plugger never uses preflag and some encoders use it more frequently. Another deviant behavior is window type. Our investigation showed that xing only uses long windows and plugger never uses short window. Probability of these features are shown in fig 3.

*Global_gain* reflects how samples have been quantized. Our investigation showed encoders are dissimilar in this aspect. Distribution of *global_gain* of different encoders are presented in fig 4a. Also, due to dissimilar behaviors of different encoders, same granule could use different number of bits. Therefore, it is expected that *part2_3_length* field in SI be quite divergent for each encoder. Our analysis showed this is the case and *part2_3_length* of no two encoders have the same distribution. In order to make the plot clearer, only distributions of four encoders are provided in fig 4b.

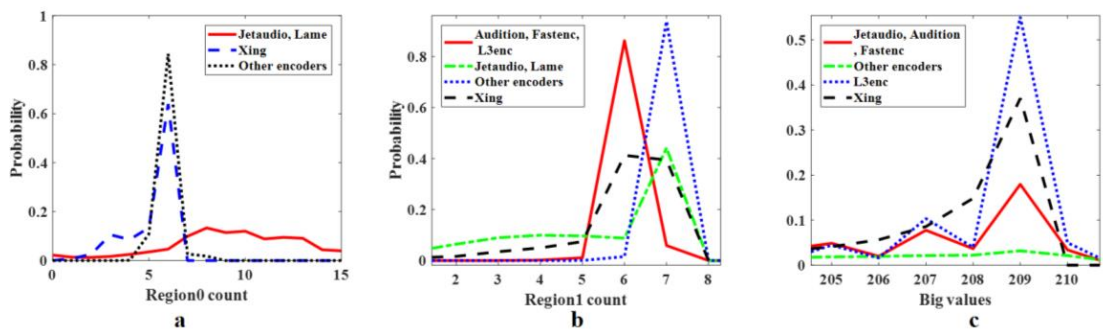

Fig 2. Distribution of bit budget   a. *region0_count*  b. *region1_count*  c. *big_values*

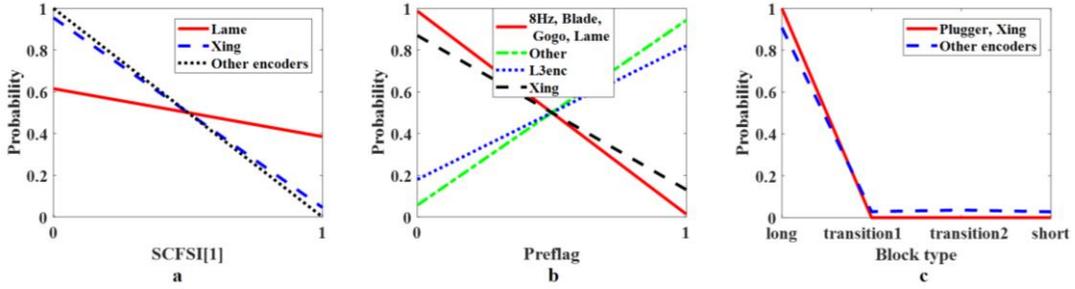

Fig 3.  Probability of using different features in encoders  a. *scfsi*  b. *preflag*  c. *block_type*

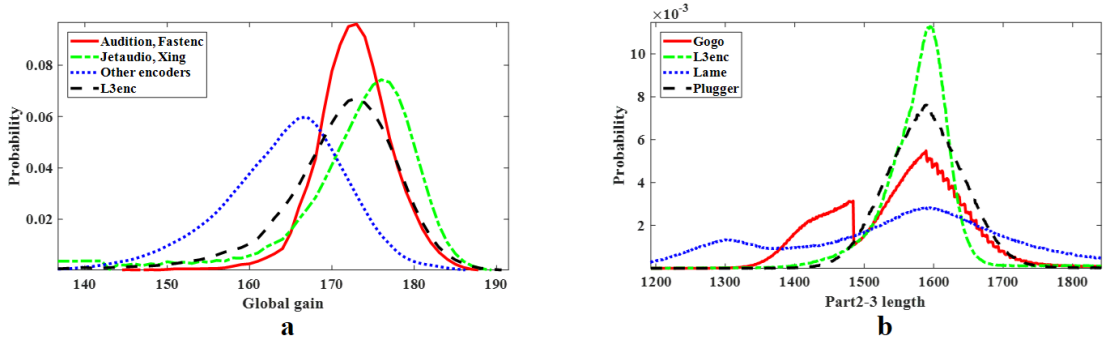

Fig 4.  Probability of different encoders    a. *global_gain*  b. *part2_3_length*

*Main_data_begin*: Our investigations showed encoders have quite different behavior in using reservoir mechanism. This, in turns will cause different distribution of *main_data_begin* field of SI. This is reflected in fig 5.a. Additionally, SI contains three fields for selecting from different Huffman tables. Our analysis showed that these fields have different distributions. Fig 5.b shows distributions of first field of *table select* for different encoders.

Our further analysis revealed that SI of first few frames have other useful information. For example, all fields of SI in the first frame of lame are always zero. *Global_gain* of both granules of first frame in plugger and *global_gain* of first granule of first frame of gogo are always 210. If there is a nonzero *scfsi* flag in the first frame, the encoder is always xing. *Rigon0_count* of first granules and second granule of first frame in plugger are always 8 and 7, respectively. *Rigon0_count* of first granules and second granule of first frame in 8Hz are always 7 and 8, respectively. Block type of first granules and second granule of first frame in plugger are always transition2 and short, respectively. Block type of granules of first frame in 8Hz are always transition1 and transition2, respectively. *Part2_3_length* of both granules of first frame in lame and plugger are always zero and *part2_3_length* of first granule of first frame in gogo is always zero. Finally, SI of first four frames of plugger always have constant values.

*2) Analysis of MDCT*

Mp3 standard does not state how frequencies higher than 16 KHz should be handled [15]. Therefore, encoders have adopted different approaches. Our analysis showed all encoders have similar characteristics for frequencies lower than 16 KHz, but after that they become quite divergent. For example, L3enc filters all components higher than 16 KHz. But, jetaudio, fastenc, audition, and xing do that after 19 KHz. Lame encode frequencies up to 20 KHz, but at the same time is has the sharpest drop in power spectrum density (PSD). Fig 6 shows PSD of different frequency regions from all encoders and for the same input signal.

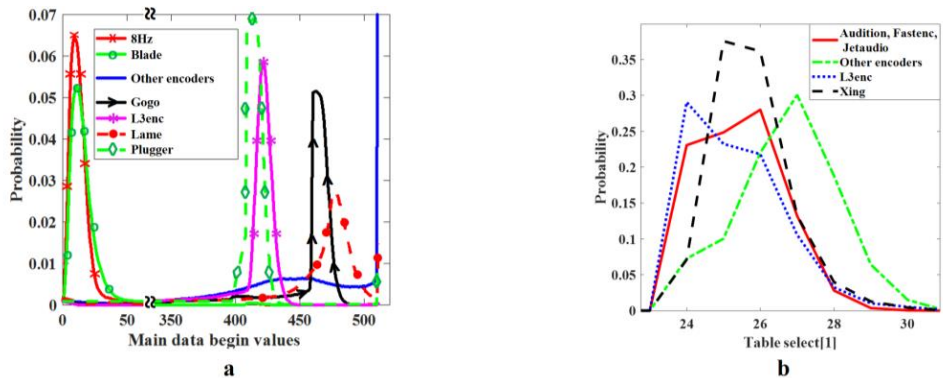

Fig 5.  Distribution of in different encoders   a. *main_data_begin*  b. *table select*

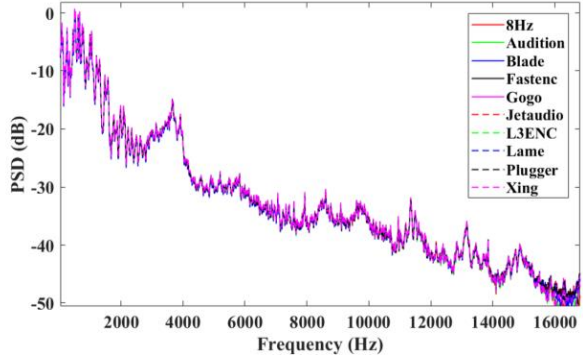 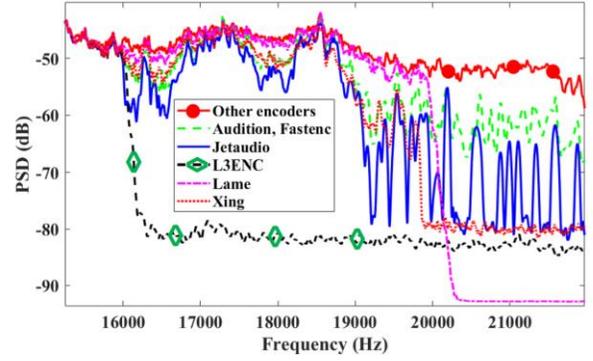

Fig 6. PSD of different encoders   a. lower than 16 KHz   b. highrt than 16 KHz

It is clear from fig 6.b that xing, audition, and fastenc have the same behavior for frequencies up to 20 Khz, but after that point there is a sharp decrease in PSD of xing.

*3) Proposed features*

We divide fields of SI into three categories. First category has one field per frame and it includes *main_data_begin* and *scfsi*. For *scfsi* all four flags were added together. Also, *main_data_begin* of some encoders were zero for the first two frames; therefore, *main_data_begin* of the first two frames were discarded prior to feature extraction. Second category has two fields per frame and each fields has significantly different distribution. This category only includes *part2_3_length*. For this category, features were extracted from each granule separately. Third category includes all the remaining fields and both granules were concatenated for feature extraction. From each measurement we extracted four different features including mean, std, min, and max. Therefore, 52 features were extracted from SI.

According to fig 6, encoders have similar behaviors for frequencies lower than 16 KHz. Doing a simple calculation shows this region starts from coefficient 1 and it ends at coefficient 418, therefore, these coefficients were discarded. The remaining coefficients from both granules were concatenated together and then they were grouped into 16 sub-bands. After summing coefficients within each sub-band, their mean, std, min, and max were calculated. Therefore, 64 features were extracted from MDCT. Block diagram of feature extraction is shown in fig 7.

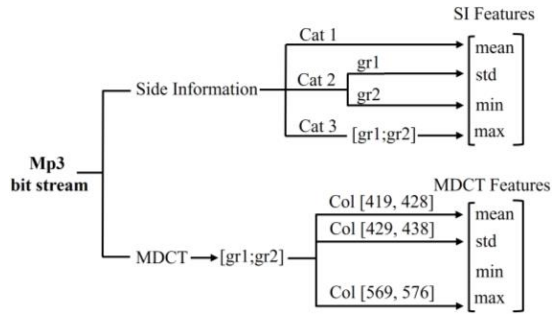

Fig 7. Proposed features for encoder classification

*B. Steganalysis of mp3stego*

Let *x* and *y* denote *global_gain* of a cover and its stego counterpart, we define noise of *global_gain* as:

$$y[m] = x[m] + n[m] \quad (1)$$

If *x* and *n* are independent random variables, then:

$$E(y) = E(x) + E(n) \quad (2)$$
$$var(y) = var(x) + var(n) \quad (3)$$

Referring to fig 8.a we see $E(n) \approx 0$ therefore:

$$E(y) \approx E(x), \quad var(y) > var(x) \quad (4)$$

It is noteworthy that analysis of mp3 algorithm predicts the same thing. The extra execution of inner loop causes an increase in *global_gain*. When compression algorithm uses higher value of quantization step, fewer bits are used and the extra bits are added to the bit reservoir. But, bit rate should be constant, so next few granules should use these extra bits which will decrease their *global_gains*. In other words, some granules of stego would have higher *global_gains* while others would have lower *global_gains*. Therefore, variance of *global_gain* in stego signal should be higher but its mean would remain the same.

Variance of *global_gain* reflects the dynamics of signal and it depends on the content of signal. Therefore, it is quite possible that variance of *global_gain* of a cover be higher than variance of *global_gain* of another stego.

$$var(x_i) > var(y_j) \quad (5)$$

In order to mitigate this, we propose a calibrated feature. Let $\mathbb{E}$ and $\mathfrak{D}$ be an estimation method and a suitable dissimilarity metric such that:

$$\mathbb{E}(x) = \tilde{x}, \mathbb{E}(y) = \tilde{y}, \quad \mathfrak{D}(x, \tilde{x}) = \varepsilon, \quad (6)$$
$$\mathfrak{D}(y, \tilde{y}) = \varepsilon, \quad \mathfrak{D}(\tilde{y}, \tilde{x}) = \varepsilon$$

where $\varepsilon$ is a small value. We define $c_y[m]$ as:

$$c_y[m] = y[m]/\tilde{y}[m] \quad (7)$$

We also define $c_x[m]$ in the same fashion. Analyzing variance of this new calibrated signal shows that:

$$var(c_x[m]) = var(x[m]/\tilde{x}[m]) \approx \varepsilon \quad (8)$$

On the other hand:

$$var(c_y[m]) = var(y[m]/\tilde{y}[m]) \not\approx \varepsilon \quad (9)$$

Therefore, variance of this calibrated signal depends mostly on embedding operation and not much on the content of signal.

The only challenge is determining the estimation method. To that end, we did some analysis on frequency properties of $x[m]$, $y[m]$, and $n[m]$. According to fig 8, $x[m]$ is a band limited signal with very low frequency components in high frequencies.

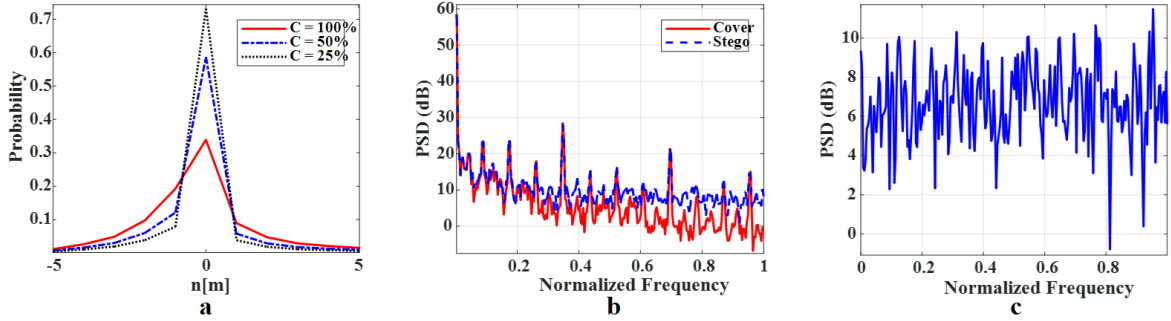

Fig 8. a.distribution of $n[m]$   b. PSD of *global_gain* of cover and stego   c. PSD of $n[m]$

On the other hand, $n[m]$ is a broad band signal and its spectrum is equally distributed over high and low frequencies. Also, it is evident that in low frequencies $x[m]$ and $y[m]$ have similar characteristics. Based on these observations we implemented $\mathbb{E}$ using a low pass filter. Furthermore, our experiments showed a simple averaging achieves quite good results. Therefore, estimation method was implemented as:

$$\tilde{x}[m] = (x[m] + x[m+1])/2 \qquad (10)$$

According to [10] difference between *global_gain* of consecutive granules is more discriminative. Also, previous audio steganalysis methods have shown that incorporating higher order statistics (HOS) can improve results of system [7, 8]. Defining $g[m]$ as:

$$g[m] = 2.(x[m] - x[m+1])/x[m] + x[m+1]) \qquad (11)$$

steganalysis features were defined as std, skewness, and kurtosis of $g[m]$.

## IV. EXPERIMENTS AND RESULTS

### A. Experiment setup

Previous works have shown that decompressed jpeg images are not good covers for jpeg embedding [19]. Considering similarities between jpeg and mp3 compressions, we addressed this concern by obtaining 27 never-compressed audio disks covering different genres of music. After cutting all tracks into 30 seconds clips, 2249 excerpts were generated. All excerpts were separately encoded with different encoders. Also, random messages were embedded into them using mp3stego algorithm at 100%, 50%, 25%, 12.5%, 6.25%, and 3.125% of maximum embedding capacity. For classification, we used SVM with 10-fold cross validation. Also, previous works have shown that normalization of features can improve results of classifications [20]. Let $\mu_k$ and $\sigma_k$ denote mean and std of feature $k$ over training set, (12) shows our procedure for feature normalization.

$$\hat{x}_k = (x_k - \mu_k)/\sigma_k \qquad (12)$$

Feature selection can reduce complexity of the system and improve its performance. For this purpose we used genetic algorithm (GA) to find the optimum set of features [7]. Our implementation used accuracy of classifier as fitness function, two-point cross over [21], and a population of 200 individuals. The algorithm was terminated if there was no improvement in the fitness function in 5 consecutive generations.

### B. Encoder classification

Efficacy of features extracted from SI and MDCT for encoder classification were tested separately for different number of features. Fig 9 shows the results.

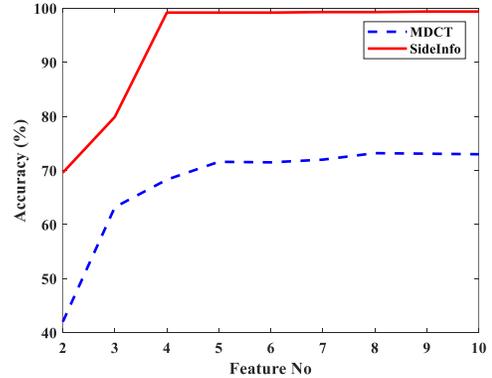

Fig 9.   Results of GA feature selection

Based on fig 9 for the rest of simulations, the best four and eight features were selected from SI and MDCT, respectively.

Encoder classification is a multi-class problem and different approaches are possible. For this purpose, performance of one against all SVM, one against one SVM, neural network (NN), and tree classifier were compared. Table I shows parameters (number of hidden layer for NN and kernel of SVM) and accuracy of each method.

| TABLE I. | ACCURACY OF DIFFERENT CLASSIFIERS | | | |
|---|---|---|---|---|
| Classifier | SI | | MDCT | |
| | Param | Acc. | Param | Acc. |
| One-One | Linear | 99.4 | guassian | 73.4 |
| One-All | guassian | 99 | guassian | 70.4 |
| NN | [9,10] | 95.7 | [9,10] | 44.1 |
| Tree | - | 98.7 | - | 72.8 |

Referring to table I, SI features with one-one SVM is the best structure. Table II shows confusion matrix of this structure.

### C. Detection of mp3stego

To show true potency of the proposed method, two different feature sets were compiled. First set only contained std, skewness, and kurtosis of (11). For the second set, we augmented first set with the best four SI features for encoder classification. Sensitivity (Se.) and specificity (Sp.) of different

TABLE II. CONFUSION MATRIX OF ONE-ONE SVM FOR DIFFERENT ENCODERS

|  | 8Hz | Audition | Blade | Fastenc | Gogo | Jetaudio | L3ENC | Lame | Plugger | Xing |
|---|---|---|---|---|---|---|---|---|---|---|
| **8Hz** | 99.8 | 0 | 0 | 0 | 0 | 0 | 0.1 | 0 | 0.1 | 0 |
| **Audition** | 0 | 100 | 0 | 0 | 0 | 0 | 0 | 0 | 0 | 0 |
| **Blade** | 6.1 | 0 | 93.8 | 0 | 0 | 0 | 0 | 0 | 0.1 | 0 |
| **Fastenc** | 0 | 0 | 0 | 100 | 0 | 0 | 0 | 0 | 0 | 0 |
| **Gogo** | 0 | 0 | 0 | 0 | 99.7 | 0 | 0 | 0 | 0.3 | 0 |
| **Jetaudio** | 0 | 0 | 0 | 0 | 0 | 100 | 0 | 0 | 0 | 0 |
| **L3ENC** | 0.3 | 0 | 0 | 0.1 | 0 | 0 | 99.5 | 0 | 0 | 0 |
| **Lame** | 0 | 0 | 0 | 0 | 0 | 0 | 0 | 100 | 0 | 0 |
| **Plugger** | 0 | 0 | 0 | 0 | 0 | 0 | 0 | 0 | 100 | 0 |
| **Xing** | 0 | 0 | 0 | 0 | 0 | 0 | 0 | 0 | 0 | 100 |

sets are reported in table III.

TABLE III. PERFORMANCE OF THE PROPOSE METHOD

|  | Feat set1 | | | | Feat set2 | |
|---|---|---|---|---|---|---|
| **Capacity** | Single encoder | | Multi encoder | | Multi encoder | |
|  | Se. | Sp. | Se. | Sp. | Se. | Sp. |
| **100** | 99.8 | 99.7 | 99.2 | 95.7 | 99.7 | 99.6 |
| **50** | 99.6 | 99.3 | 98.4 | 93.1 | 99.5 | 99 |
| **25** | 98.9 | 98.6 | 96.5 | 88.1 | 99 | 97.4 |
| **12.5** | 97.8 | 96.8 | 95.7 | 81.4 | 98.4 | 94.9 |
| **6.25** | 96 | 95.7 | 95.9 | 74.6 | 97.9 | 93.6 |
| **3.12** | 96.3 | 95.4 | 95.5 | 76.1 | 97.8 | 94.5 |

Fig 10 shows receiver operating characteristic (ROC) of different scenarios for mp3stego at embedding capacity of 3.12%.

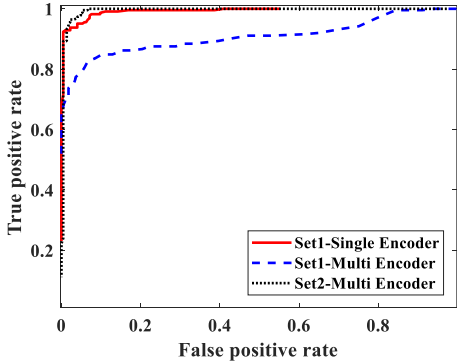

Fig 10. ROC of different scenarios

Referring to table III and fig 10, in set 1, when multiple encoders are employed performance of the system is degraded considerably. This is due to the fact that distribution of *global_gain* in some encoders are more similar to mp3stego than 8Hz encoder. Therefore, these samples would be misclassified as stego, hence the low value of specificity. On the other hand, when features for classifying encoders are added, this problem is solved and a big improvement is observed.

## V. DISCUSSION AND CONCLUSION

Performance of the proposed method is compared with some of previous works in both single encoder and multi encoder scenarios. For this purpose methods of block length (BL) [9], bit reservoir (BR) [11], quantization steps (QS) [10], markov transition probability of MDCT (MDCT) [12] and markov transition probability of difference of absolute values of MDCT (DAMDCT) [13] were implemented. DAMDCT method originally had 338 features. We used procedure of the original paper for feature selection and selected the best 34 features. We used the following parameters for this purpose.

$$\varepsilon_{inter\_mean} = 0.01, \quad \varepsilon_{inter\_std} = 0.05$$
$$\varepsilon_{intra\_mean} = 0.015, \quad \varepsilon_{intra\_std} = 0.05$$

Our analysis showed DAMDCT produces the same results regardless of embedding capacity, which is counter intuitive. We expect sensitivity and specificity to reduce for lower embedding capacities. Investigating mp3stgo algorithm showed that, frames are not selected quite randomly. Therefore, at low capacities only the first few frames are modified. Also, DAMDCT method only uses the first 200 granules for feature extraction. Apparently, if frames are selected randomly, this method may produce quite different results. To estimate the true potency of this method, we followed the same procedure but features were extracted from all frames of mp3. For feature selection we used the same procedure with following parameters.

$$\varepsilon_{inter\_mean} = 0.0015, \quad \varepsilon_{inter\_std} = 0.02$$
$$\varepsilon_{intra\_mean} = 0.003, \quad \varepsilon_{intra\_std} = 0.01$$

Results of this modified method are denoted by DAMDCT II in the table IV.

TABLE IV. COMPARISON WITH PREVIOUS WORKS

| Method | Feat No. | Single encoder capacity | | | | | | Multi encoder capacity | | | | | |
|---|---|---|---|---|---|---|---|---|---|---|---|---|---|
|  |  | 100 | 50 | 25 | 12.5 | 6.25 | 3.12 | 100 | 50 | 25 | 12.5 | 6.25 | 3.12 |
| **BL** | 1 | 99.9 | 99.3 | 97.5 | 94.6 | 91.7 | 91.3 | 80.9 | 76.1 | 79.6 | 81.6 | 80.1 | 82.4 |
| **BR** | 1 | 80.7 | 80.9 | 81.6 | 81.6 | 81.1 | 81.3 | 87.9 | 87.7 | 87.7 | 87.7 | 87.5 | 87.5 |
| **QS** | 1 | 99.7 | 99.1 | 97.6 | 94.4 | 91.7 | 91.2 | 91.4 | 85.6 | 79.6 | 79. | 78.9 | 79 |
| **DAMDCT** | 34 | 97.6 | 97.8 | 98.2 | 97.7 | 98 | 98 | 93.1 | 93.2 | 94.1 | 93.1 | 93.6 | 94 |
| **DAMDCT II** | 56 | 99.8 | 98.3 | 93.7 | 86.9 | 81.7 | 81.6 | 96.6 | 92.3 | 84.1 | 74.5 | 70.2 | 70 |
| **MDCT** | 48 | 97.9 | 93.9 | 84.3 | 75.9 | 70.7 | 70.4 | 93. | 85.6 | 79.2 | 74.5 | 72.9 | 73.7 |
| **Proposed** | 7 | 99.7 | 99.2 | 98.7 | 97.3 | 95.9 | 96 | 99.6 | 99.1 | 98.2 | 96.5 | 95.9 | 95.7 |

MDCT method originally had 294 features. We used two stage feature selection proposed in the original paper and achieved the best result for 48 features. Table IV compares accuracy of the proposed method with previous works.

Referring to table IV we see a big difference between performance of DAMDCT and DAMDCT II. So, if embedding is done randomly performance of this method would degrade considerably. Furthermore, in multi encoder scenario which is more realistic case, it is evident that performance of previous works degrades a lot. On the other hand, not only the proposed method has a better performance in the single encoder scenario, but also its performance remains the same in the multi encoder case.

This paper investigated dissimilarities between different mp3 encoders and showed that these subtle differences are present in all fields of SI and even their MDCTs. We showed that only four features from SI is enough for almost perfect classification. Also, it was shown that differences between spectra of encoders can achieve accuracy of 73.4%. Based on these observation, we hypothesized that accuracy of existing methods would not be satisfactory in realistic settings where different encoders are present. Our analysis showed decrease in accuracy of previous works could be as high as 23.2%. In order to solve this issue, we augmented steganalysis features with encoder classification features. Furthermore, for steganalysis features we showed that *global_gain* was a band limited signal and effects of steganography noise were more prominent in its high frequency regions. We used these observations and applied a low pass filter for calibrating features extracted from *global_gain*. Simulation results showed that accuracy of the proposed method remains the same in single and multi-encoder scenarios and it achieves very high level of accuracy even at low embedding rates.